\documentclass[aps,nofootinbib,a4paper,superscriptaddress,twocolumn]{revtex4-2}
\usepackage{tensor}
\usepackage{graphicx}
\graphicspath{ {images/} }
\usepackage[utf8]{inputenc}
\usepackage{amsmath}
\usepackage{mathtools}
\usepackage{amssymb}
\usepackage{enumerate}
\usepackage{subfigure}
\usepackage{tabularx}
\usepackage{soul}
\usepackage{lipsum} 
\usepackage{float} 
\usepackage[colorlinks=true, pdfstartview=FitV, linkcolor=blue, citecolor=red, urlcolor=black, breaklinks=true]{hyperref}
\usepackage{orcidlink}
\newcommand{\cmmnt}[1]{}
\newcommand{\be}{\begin{equation}}
\newcommand{\ee}{\end{equation}}
\newcommand{\ben}{\begin{eqnarray}}
\newcommand{\een}{\end{eqnarray}}
\newcommand{\bes}{\begin{subequations}}
	\newcommand{\ees}{\end{subequations}}
\def\bal#1\eal{\begin{align}#1\end{align}}

\newcommand{\LL}{{\mathcal L}}

\newcommand{\qt}[1]{``#1''}
\newcommand{\pd}[2]{\ensuremath{\frac{\partial#1}{\partial#2}}}

\begin{document}
	\title{Geometrically constrained multi-kink configurations in generalized impurity-doped field theories}
	
\author{D. Bazeia\,\orcidlink{0000-0003-1335-3705}}\email[]{ bazeia@fisica.ufpb.br}\affiliation{Departamento de F\'\i sica, Universidade Federal da Para\'\i ba, 58051-970 Jo\~ao Pessoa, PB, Brazil}
	\author{M. A. Liao,\orcidlink{0000-0001-9720-2079}}\email[]{matheusalvesliao@gmail.com}\affiliation{Departamento de F\'\i sica, Universidade Federal da Para\'\i ba, 58051-970 Jo\~ao Pessoa, PB, Brazil}
\author{M.A. Marques\,\orcidlink{0000-0001-7022-5502}}
\email[]{mam@fisica.ufpb.br}\affiliation{Departamento de F\'\i sica, Universidade Federal da Para\'\i ba, 58051-970 Jo\~ao Pessoa, PB, Brazil}

	\begin{abstract}
		 This short communication investigates impurity coupling in generalized field theories where scalar coupling  is introduced directly at the level of the kinetic and gradient contributions of the energy. We show that the fundamental aspects of the original theory, which has been previously investigated in the impurity-free setting, can be extended to the inhomogeneous scenario. In particular, an interpretation in terms of geometrically-constrained effective one-field theories with impurities is possible in the separable case. We show that BPS multi-kink configurations are possible in the model, as well as in the usual half-BPS scalar theories.
	\end{abstract}
	
	\date{\today}
	\maketitle

%\section{Introduction}\label{Intro}

In recent years, several papers have explored the role of impurities in scalar field theories~\cite{malomed,KinkI, KinkII, KinkIII,AdamI, AdamII, KinkKink, BLMPLB,BLM24, BMM24}. In these works, homogeneous field theories are coupled to bounded, and usually localized, functions representing inhomogeneities of space, thus breaking translational invariance. Such frameworks are particularly useful as effective theories, where translational symmetry may often be broken due to the presence of external fields, particles, or other background effects. These results have been extended to two-field theories in a recent work~\cite{2fieldimp}, where a canonical Lagrangian was deformed to account for coupling with two impurity functions in such a way as to preserve part of the BPS sectors of the theory. We now further extend this framework to include a nonstandard derivative coupling of the form introduced in Ref.~\cite{BLM2020}, and show that the main features of that work, including the geometrically constrained kink-like solutions that emerge when superpotentials are separable, can be extended to account for impurity doping. The model introduced in Ref.~\cite{BLM2020} has been revisited in many subsequent works by several authors. These works include, but are not limited to, references~\cite{L1,L2,L3,L4,L5,L6,daHora, Lima}. Therefore, extension of these results to the inhomogeneous scenario is a worthwhile endeavor. 

Among the most interesting new possibilities that ensue from impurity-doping is the presence of BPS-saturating  multi-kink solutions. These solutions correspond to one-field configurations with integer topological charge greater than unity, meaning that they interpolate between three or more vacua. In the homogeneous scenario, this feature is usually impossible for time-independent solutions~\cite{rajaraman}. In Ref.~\cite{KinkKink}, it has been shown that this limitation can be circumvented in the presence of impurities, and that double-kink configurations can be found in a doped Sine-Gordon model with BPS equation $\phi'=2\sin(\phi/2)s(x)$, where $s(x)$ has a pole at some point $x_0$ inside the domain, thus forcing the field to reach a vacuum value at this point. Here we show that multi-kink configurations can be obtained both in the one-field theory introduced in~\cite{AdamI} and in a novel impurity-doped version of the models introduced in~\cite{BLM2020}, which possesses a multiplying factor $1/P(\phi,\chi)$. This factor gives rise to a removable singularity when $P(\phi,\chi)$ has zeros. It will  be shown that this construction can be consistently used to generate multi-kink configurations.

%\section{Generalized model coupled to impurity}

We work in two-dimensional Minkowski spacetime with  metric signature $(+,-)$ and Lagrangian density 
\begin{equation}\label{Lagrangian}
	\begin{split}
		\mathcal{L} =& \frac{P}{2}\left[\partial_{\mu}\phi\partial^{\mu}\phi + 2\phi'\sigma(x) -\left(\frac{W_{\phi}}{P}+\sigma(x)\right)^2\right] \\
		& \ \ \ \ \ + \frac{1}{2}\partial_{\mu}\chi\partial^{\mu}\chi -\frac{1}{2}W_{\chi}^2,
	\end{split}
\end{equation}
where $\sigma(x)$ is a bounded, differentiable function that models the strength of the impurity at each point, $P=P(\phi,\chi)\geq 0$ is at least almost everywhere differentiable and  $W(\phi,\chi)$ is an auxiliary function playing the role of a superpotential. We assume that $\sigma(x)$ tends to zero as  $x\to\pm\infty$, thus representing an impurity that is only non-negligible inside some neighborhood of the origin, although the above Lagrangian also makes mathematical sense if $\sigma(x)$ is not localized. Furthermore, we assume $P(\phi,\chi)$ tends to a nonzero constant as $(\phi,\chi)$ approaches a minimum of $V_{0}(\phi,\chi)$, where 
\begin{equation}
V_{0}(\phi,\chi)=\frac{1}{2}\left(\frac{W_{\phi}^2}{P}+W_{\chi}^2\right).
\end{equation}
The above Lagrangian reduces to the one used in Ref.~\cite{BLM2020} when $\sigma=0$, which happens by hypothesis at large distances from the the origin.The particular coupling is inspired by the work of Refs.~\cite{AdamI, AdamII} in the one-field scenario. In Ref.~\cite{A2} the authors show that this kind of impurity coupling can be understood as the effect of an infinitely heavy frozen kink in a two-field theory.

Our assumptions imply that acceptable boundary conditions for finite energy configurations - and thus the topology of the theory - are inherited from the homogeneous scenario. Moreover, the kinetic, gradient and potential contributions are reduced to that of a canonical theory when $P$ equals a constant. In the separable models considered below, this limit is achieved if and only if $\chi$ approaches a \qt{vacuum value}, i.e., a zero of $W_{\chi}$. The field equations of the theory are 
\begin{subequations}\label{EulerLagrange}
\begin{equation}\label{EulerLagrangeA}
	\begin{split}
	&\partial_\mu \left[ P(\phi,\chi) \left( \partial^\mu\phi + \delta^{x\mu}\sigma(x) \right) \right] +V_{0}(\phi,\chi)_{\phi} \\ &  \quad -\frac{P_\phi(\phi,\chi)}{2}\left(\partial_\mu\phi\partial^\mu\phi + 2\phi'\sigma(x)-\sigma(x)^2 \right)\\
& \quad \quad \quad  +\sigma(x) W_{\phi\phi}(\phi,\chi)  =0
	\end{split}
\end{equation}
and 
\begin{equation}\label{EulerLagrangeB}
	\begin{split}
		& \partial_\mu \partial^\mu\chi  +V_{0}(\phi,\chi)_{\chi}  +\sigma(x) W_{\phi\chi}(\phi,\chi) \  \\ & - \frac{P_\chi(\phi,\chi)}{2}\left(\partial_\mu\phi\partial^\mu\phi + 2\phi'\sigma(x)-\sigma(x)^2 \right) =0
	\end{split}
\end{equation}
\end{subequations}

The above equations must be accompanied by boundary conditions derived from the assumption of finiteness of the energy. Compatibility with this condition restricts the allowed values of $(\phi,\chi)$ at $x\to\pm\infty$ to a discrete set $\mathcal{M}$, which must have more than one element if topological defects are to be found in this theory. Since impurities are assumed to be localized in this work, $\mathcal{M}$ is identical to  the zero set of $V_{0}$.

The coupling between fields and impurities  is not arbitrary: it has been chosen to allow for a Bogomol'nyi bound~\cite{bogo}. Indeed, we can complete squares in the standard energy density $\rho= \pd{\LL}{\dot{\phi}}\dot{\phi}+\pd{\LL}{\dot{\chi}}\dot{\chi} -\LL $ to demonstrate that, if the boundary conditions give rise to a topologically nontrivial space, the energy functional is subject to the bound 
\begin{equation}\label{BgmBound}
	E\geq \Delta W\equiv W(\phi(\infty),\chi(\infty))-W(\phi(-\infty),\chi(-\infty)),
\end{equation}
 with saturation if, and only if, the solution is static and
 \begin{subequations}\label{BPSI}
	\begin{align}
		\phi'(x)&=\frac{W_{\phi}(\phi,\chi)}{P(\phi,\chi)}+\sigma(x),  \label{FOA}\\
		\chi'(x)&=W_{\chi}(\phi,\chi).\label{FOB}
	\end{align} 
\end{subequations}

Although the framework here developed is more general, this work shall mostly deal with separable models, which are obtained when the functions in the Lagrangian satisfy the conditions 
\begin{align}\label{GeometricallyConstrained}
	W_{\chi\phi}=W_{\phi\chi}=0 && P=P(\chi).
\end{align}

In this case, the $\chi$ equation can be solved independently and the result inserted into~\eqref{FOA} to obtain an ordinary differential equation for $\phi$. The usefulness of this class of models lies not only in the relative simplicity of the field equations but also, and more importantly, in the fact that these solutions are equivalent to kink profiles of a one-field theory in a geometry generated by $\chi(x)$. Indeed, the fact that $\chi(x)$ is independent of $\phi(x)$ in the BPS limit implies that $P(\chi(x))$ acts as a one-dimensional multiplying factor scaling $\phi'(x)$, thus acting as a geometric constriction for $\phi$. One can  locally define a variable $\xi(x)$ by the equation $d\xi/dx=P(\chi(x))^{-1}$, which we can use to write~\eqref{FOA} in the form:
\begin{equation}\label{BPSCsi}
\frac{d\phi(\xi)}{d\xi}= W_{\phi}(\phi(\xi)) + \sigma(\xi).
\end{equation}
 
These solutions are therefore related to the ones investigated in ~\cite{AdamI, AdamII} by the transformations $x\leftrightarrow \xi$, thus extending the geometric framework of~\cite{BLM2020} to the impurity-doped scenario. Indeed,~\eqref{BPSCsi} is identical to the BPS equation of a model with effective Lagrangian 
 \begin{equation}\label{effectiveLag}
 	\LL_{\text{eff}}=\frac{1}{2}\left(\partial_{\tilde\mu}\phi\partial^{\tilde\mu}\phi - W_{\phi}^2\right) - \left(W_{\phi}-\partial_{\xi}\phi\right)\tilde\sigma(\xi) +\frac{\tilde\sigma^2(\xi)}{2}
 \end{equation}
  in a geometry specified by the metric $\tilde g_{\mu\nu}=dt^2-d\xi^2$ once the identifications $d\xi=(P(x))^{-1}dx$ and $P\sigma\leftrightarrow\tilde\sigma$ are made. In this sense, we are able to map our $\phi$ profiles into solutions of a one-field theory with Lagrangian $\LL_{\text{eff}}$, and may thus treat the localized solutions encountered therein as kinks, despite the fact that we were originally dealing with two-field theories. \textcolor{black}{This is an useful analogy that stems from the simplification enabled by BPS saturation.} The energy density of the effective one-field system can be calculated in the usual way from the above Lagrangian. After returning to the original $x$ coordinate and using the BPS equation~\eqref{BPSCsi}, one can write the resulting static energy in the form:
  \begin{equation}\label{rhoeff}
 	\rho_{\text{eff}}=\underbracket{P(x)\frac{\phi'^2}{2} + \frac{W_{\phi}^2}{2P(x)}}_{\rho_{(\phi)}} + \underbracket{\frac{\tilde\sigma(x)}{2}\left(W_{\phi}-P(x)\phi'\right)}_{\rho_{c}},
  \end{equation}
where the quantities in underbrackets discriminate the energy density in the \qt{intrinsic} portion $\rho_{(\phi)}$, associated only to the field's gradient and self-interaction energies, and the portion $\rho_{c}$ which integrates to the coupling energy between scalar field and impurity plus the background energy associated to $\tilde\sigma^2/2$. Note that, using the BPS equation, one could alternatively write $\rho_c=-\tilde\sigma^2/2$, but we choose the form above due to the clearer physical interpretation of the terms. The integral of $\rho_{(\phi)}$ is the number one might reasonably associate with the energy of the kinks themselves, while $\rho_{\text{eff}}$ accounts for the energy of the entire defect-impurity system.
  
The topology of the theory is determined by the topological group generated by the allowed maps between spatial infinity and $\mathcal{M}$, and thus depends on the asymptotic values of both fields. By the logic outlined above, it nevertheless makes sense, when dealing with BPS/near-BPS solutions, to  attribute a \qt{topological charge} to $\phi$, which can be rigorously defined by identification with the topological charge of the system generated by~\eqref{effectiveLag}, into which $\phi(x)$ is mapped by~\eqref{FOA}. We may thus write~\cite{rajaraman, manton}
\begin{equation}\label{topologicalcharge}
	Q_{T}=\frac{1}{\Delta v}\int_{-\infty}^{\infty}\partial_{x}\phi \ dx
\end{equation}
where $\Delta v$ is an arbitrary normalization factor. If, as is usually the case, the minima of the potential are equally spaced, $\Delta v$ may be identified with their difference. $Q_{T}$ is hence an integer amounting to a summation of all  (anti)kinks, each of which corresponds to (negative) positive unit charge. It must be understood, of course, that these identifications are valid only as long as $\chi$ is held fixed as a solution of~\eqref{FOB}, which is reasonable if the back-reaction caused by $\phi$ in the event of a small perturbation is weak enough not to cause any appreciable movement in $\chi$.

We now turn our attention to static solutions that may be interpreted as multi-kink configurations. These are obtained when a scalar field interpolates between three or more distinct vacua, so that, at least for sufficiently separated defects, one can identify more than one independent kink~\cite{manton}. Such solutions are usually not possible in the canonical one-field theory, as can be seen through a mechanical analogy~\cite{rajaraman} achieved through the relations $\phi\leftrightarrow q(t)$ $V(\phi)\leftrightarrow -U(q(t))$. The static equation in the canonical one-field model is thus mapped into a mechanical equation of the form $\dot{q}(t)=\sqrt{2U(q)}$, which implies that the analogue particle reaches the vacuum with zero momentum and acceleration~\cite{rajaraman}. When an impurity is added, this argument ceases to hold, as the analogous mechanical system becomes non-conservative.

Before moving to explicit examples of this feature, let us turn our attention to a subtlety that has not been properly addressed in Ref.~\cite{BLM2020} or any of the works derived from it. The Lagrangian~\eqref{Lagrangian}, and thus everything stated above, is perfectly valid as long as $P(\phi,\chi)$ is bounded and nonzero. However, $\LL$ becomes singular whenever one of these assumptions is violated, and thus the equations of motion are not everywhere defined in the classical sense. Nevertheless, it is still possible to obtain valid solutions in these cases as long as all terms in the field equations are sufficiently regular. This is in fact the case of the models presented in Ref.~\cite{BLM2020}, which amount to the $\sigma(x)=0$ case of Eqs.~\eqref{BPSI}. These configurations possess plateaus at poles of $P$, as $\phi'$ must vanish at such points for any bounded $W_{\phi}$. \textcolor{black}{The expression $\partial_{x}\left(P\partial_{x}\phi\right)\to 0$ has a well-defined limit, thus regularizing the Lagrangian and energy density on-shell. If, on the other hand, $P(\phi,\chi)$ possesses an isolated zero instead of a pole, Eq.~\eqref{BPSI} implies that $\phi'$ can only be regular if $W_{\phi}$ vanishes with a zero of equal or higher order. } 

The situation described above can be formalized - and in fact greatly generalized \textcolor{black}{at the level of the second-order equations,} through the notion of \emph{weak solutions}. Let the fields belong to Sobolev spaces $H_0^1(\mathbb{R})$ - meaning that they possess at least one weak derivative defined with respect to a suitable $L^2$-norm~\cite{SobolevSpaces} - and consider the static case of equations~\eqref{EulerLagrange}. We now multiply these equations by smooth test functions with compact support and, after integration by parts, find:
\begin{subequations}\label{weakformulation}
\begin{equation}
	\begin{split}
		& \hspace{40pt}\int_{\mathbb R} P\,(\phi'-\sigma(x))g'(x)\,dx \quad + \\
		& \quad \int_\mathbb R \Bigg[ 
		W_{\phi\phi}\left(\frac{W_\phi}{P}+\sigma(x)\right)
		+ W_{\chi\phi}W_\chi\\
		& \quad \quad + \frac{P_\phi}{2}\left((\phi'-\sigma(x))^2-\frac{W_\phi^2}{P^2}\right)
		\Bigg]g(x) \, dx=0,
	\end{split}
\end{equation}
and 
\begin{equation}
	\begin{split}
		& \int_\mathbb R \chi'\,h'(x)\,dx + \int_\mathbb R \Bigg[
		W_{\phi\chi}\Big(\frac{W_\phi}{P}+\sigma(x)\Big) -\frac{P_{\chi}W_\phi^2}{2P^2}
		\\
		&
		+ W_{\chi\chi}W_\chi + \frac{P_\chi}{2}(\phi'-\sigma(x))^2
		\Bigg] h(x) \, dx =0.
	\end{split}
\end{equation}
\end{subequations}
for all test functions $g(x),h(x) \in H_0^1(\mathbb R)$. The above equations may be well-defined even if $P$ or $1/P$ have isolated singularities. If this is the case for an arbitrary pair of test functions  $g(x),h(x)$, then $(\phi,\chi)$ is said to be a weak solution of the Euler-Lagrange equations. For a deeper explanation of weak solutions and formal definitions of Sobolev spaces and weak derivatives, the reader is directed to~\cite{SobolevSpaces, PDEs}. For our purposes, it suffices to state that solutions shall be considered acceptable as long as the above integral equations hold. Thus, we need not require the differential equations to be solved at every point in $\mathbb R$, as long as the above construction makes sense.

When $P(\phi,\chi)$ has isolated zeros or poles, the formulation above implies that the  weak solution is identical to a strong one outside the discrete set of points specified by these singularities. Using Sobolev embedding theorems~\cite{SobolevSpaces}, one may show that, in the one-dimensional case considered, the solutions may be taken as absolutely continuous, and are hence obtainable by solving the equation in the singularity-free intervals and patching together the results by imposing continuity at the pole. Another method, which is specially useful in numerical applications \textcolor{black}{when $P$ has zeros, consists of replacing $P$ with $P_{\epsilon}=P + \epsilon$, $0<\epsilon <<1$, which, being strictly positive, does not give rise to any singularity. Thus, to each $\epsilon$ there corresponds a strong solution $(\phi_{\epsilon}, \chi_{\epsilon})$.} Since the energy functional and, by continuity, $W_{\phi}(\phi_{\epsilon})/(P + \epsilon)$ are bounded, it can be shown that this sequence converges to a weak solution $(\phi(x),\chi(x))$ as $\epsilon\to 0$. \textcolor{black}{If $P$ has poles instead of zeros, a similar regularization scheme can be developed by replacing $1/P$ with $1/\left(P + \epsilon\right)$, $0\leq\epsilon <<1$\cmmnt{~\footnote[2]{\textcolor{black}{Note that in this case the BPS equations are everywhere defined even when $\epsilon=0$, since the right-hand side of Eq.~\eqref{FOA} simply equals $\sigma(x)$ in this case. This means that the corresponding configurations are {strong} solutions of the first-order equations. However, the \emph{second-order} equations are also undefined at poles of $P$ due to products involving $P$ and its derivatives, so that, at the level of the Euler-Lagrange equations, zeros and poles of $P$ are on equal footing. The fact that only the latter amount to weak solutions of the BPS equations follows from an ambiguity in the Bogomol'nyi procedure, as one can just as well complete squares in a way that implies the first-order equations $P\phi'=W_{\phi}$. Thus, whether or not the BPS equations are pointwise solved everywhere depends on which of the physically equivalent Bogomol'nyi formulations is chosen. }}}.}This gives a theoretical tool that can be used to solve Eqs.~\eqref{BPSI} both formally and numerically, as regularization makes some numerical computations significantly easier. More importantly, this ensures that all products involving $P$ or $1/P$ in~\eqref{Lagrangian} are finite on-shell, so the singularities are removable and cause no concern from a physical standpoint. \textcolor{black}{This regularization scheme is particularly useful in solving the stability equations, where the many removable singularities related to zeros or poles of $P$ can lead to stiff systems. This analysis will be discussed in detail elsewhere.}

Note that the considerations above imply that $W_{\phi}$ must vanish at a zero of $P(\chi)$, but the function is not forced to remain at a zero at either side of the singularity, as the contribution from $\sigma(x)$, and thus $\phi'$, remains nonzero, so multi-kink configurations are now possible. This situation is similar to the mechanism used to generate double-kink solutions in Ref.~\cite{KinkKink}, where a one-field theory with BPS equation $\phi'=W_{\phi}s(x)$ is considered. In the aforementioned reference it is shown that multi-kink configurations can appear in these systems if $s(x)$ has poles, as continuity forces $W_{\phi}=0$ at this point, while the behavior at lateral neighborhoods of the pole need not be trivial. In our case, the pole from $P(\chi)$ plays a similar role as that of $s(x)$ in Ref.~\cite{KinkKink}: it serves to ensure that a minimum of the potential is reached at a finite $x_0$. 

\textcolor{black}{As we will later demonstrate, poles are not strictly necessary to construct multi-kink solutions, which can also be found in the the half-BPS theories of Refs.~\cite{AdamI, AdamII}. Nevertheless, the models with regularizable singularities are simple and have some interesting analytic properties, so we shall start with them. To this end, consider  the model specified by the functions:}
\begin{subequations}
\begin{align}
	&\hspace{80pt}P(\chi)=\chi^{2},   \\
	&W(\phi,\chi)
	=
	\frac{1-\alpha}{2}\,\phi
	\;+\;
	\frac{1+\alpha}{4}\,\sin(2\phi) + \chi - \frac{\chi^{3}}{3},
\end{align}
\end{subequations}
where $\alpha$ is a nonnegative real constant. To fully specify the problem, let us consider an impurity function $\sigma(x)=Axe^{-(x-\bar{x})^2} H(x)$,  where $\bar{x}$ is a real constant and $H(x)$ is the Heaviside step function, defined as  $H(x)=1$ for $x\geq 0$ and  $H(x)=0$ for $x< 0$. Under these conditions, the BPS equation for $\chi$ is the usual $\chi^4$ equation with solution $\chi(x)=\tanh(x-x_0)$ while~\eqref{FOA} becomes
\begin{equation}\label{Ex2}
\phi'(x)=\frac{\cos^2(\phi)-\alpha\sin^2(\phi)}{\tanh^2(x-x_0)}+Axe^{-(x-\bar{x})^2} H(x), 
\end{equation}
where the $\chi^2(x)$ factor in the denominator has been substituted. Since $\sigma(x)$ is a localized function, the boundary conditions satisfied by finite-energy configurations must be such that the fields at infinity take values in the set:
\cmmnt{
 %\begin{equation}
 %\mathcal{M}=\left\{(\phi,\chi): \phi=\pm\arctan\!\left(\frac{1}{\sqrt{\alpha}}\right) + n\pi , \chi=\pm 1 \right\}
 %\end{equation}

  \begin{equation}
 	\begin{split}
 		\mathcal{M}= &\left\{ (\phi,\chi): \phi=\pm\arctan\!\left(\sqrt{\alpha^{-1}}\right) + n\pi, \right. \\
 		&\left.  \hspace{40pt}\chi=\pm 1 \right\},
 	\end{split} 
 \end{equation}}
 \begin{equation}\label{vacua}
\mathcal{M}= \left\{ (\phi,\chi)=\left(\pm\arctan\!\left(\sqrt{\frac{1}{\alpha}}\right) + n\pi, \ \pm 1\right) \right\},
 \end{equation}
  \noindent where $n\in\mathbb{Z}$.  Any solution that interpolates between  minima of the form $(\phi^-,-1)$ and $(\phi^+,1)$ is such that $\chi(x)=\tanh(x-x_0)$, since $\chi$ doesn't depend on the asymptotic values of $\phi$. Using either the $\epsilon$ construction discussed above or the stated continuity requirement for BPS solutions we can show that $\phi(x)$ can be specified by solving~\eqref{FOA} separately  in $(-\infty,x_{0})$ and $(x_{0}, \infty)$ and matching the values at $x_{0}^{-}$ and $x_{0}^{+}$. Naive integration of the $A=0$ case of~\eqref{FOA} leads to a global solution with discontinuous character, which, as remarked, would be incompatible with our weak formulation and with~\eqref{BgmBound}, but we may use the piecewise construction mentioned above to find continuous BPS solutions in this topological sector. The technique amounts to solving the BPS equations in the regions to the left and right of $x_0$ and then solving the initial value problem implied by the condition $\phi(x_{0}^{-})=\phi(x_{0}^{+})$. This leads to two possibilities, namely,
\cmmnt{
\begin{equation}
	\phi_{0}(x)=\arctan \left[ \frac{1}{\sqrt{\alpha}} \tanh \left( \sqrt{\alpha} (u - \coth u + c), \right) \right]
\end{equation}
for }
\begin{equation}\label{Solhom1}
\phi_{0}(x)=\begin{cases}
	\arctan \left[ \frac{\tanh\left(\sqrt{\alpha}(\xi(x)-\xi_0)\right)}{\sqrt{\alpha}} \right]  & \text{for} \ x < x_0 \\
	\hspace{40pt} \arctan\!\left(1/\sqrt{\alpha}\right)  & \text{for} \ x \ge x_0
\end{cases}
\end{equation}
and
\begin{equation}\label{Solhom2}
	\phi_{0}(x)=\begin{cases}
		\hspace{30pt} -\arctan\!\left(1/\sqrt{\alpha}\right)  & \text{for } x \leq x_0 \\
	\arctan \left[ \frac{\tanh\left(\sqrt{\alpha}(\xi(x)-\xi_0)\right)}{\sqrt{\alpha}} \right]  & \text{for} \ x > x_0
	\end{cases}
\end{equation}
\noindent where $\xi\equiv x-x_0-\coth(x-x_0) -c$, with $c$ an arbitrary real constant. Each combination of $x_0$ and $c$ specifies a solution, so that the moduli space generated by Eqs.~\eqref{BPSI} can be parameterized by these two constants in the homogeneous limit. Both of these solutions present a half-compact profile, as the singularity at $x=x_0$ forces $\left.W_{\phi}\right|_{x=x_0}=0$, thus forcing $\phi$ to attain a vacuum value at this point. This is a general feature of the $\sigma(x)=0$ case: any weak solution saturating~\eqref{BgmBound} must be either a half-compacton or compacton~\cite{Compactons} (if $P(\chi(x))$ has multiple zeros in the domain) kink in the homogeneous scenario, as $W_{\phi}$ and all derivatives of $\phi$ must vanish at these points. 

When impurities are added, $W_{\phi}$ must still vanish at the zeros of $P$, but compactification does not occur since there remains a nonzero contribution from $\sigma(x)$. Defining $\delta^{\pm}(x)\equiv \phi(x)\mp \arctan\!\left(1/\sqrt{\alpha}\right)$, thus accounting for the two solutions above, and linearizing~\eqref{FOA} in a neighburhood  $U_{0}$ of $x_0$, we find
\begin{equation}
	\delta^{\pm}(x)=c_1e^{\pm\frac{2\sqrt{\alpha}}{x}} \pm \frac{Ax^3}{2\sqrt{\alpha}}e^{-\bar{x}^2},
\end{equation} 
where the first contribution comes from the homogeneous solution of the linearized equation (and is hence identical to the result of the impurity-free case) while the second one is the particular solution compatible with the impurity and initial conditions. Since $e^{\pm\frac{2\sqrt{\alpha}}{x}}$ blows up as $x\to 0^{+}$, the upper sign solution must be such that $c_1=0$, which is consistent with our earlier claim of compactness in the $A=0$ case. On the other hand, the homogeneous contribution in $\delta^{-}(x)$ need not vanish entirely, but its exponentially fast decay within a right neigbourhood of the origin is dominated by the power-law contribution from the particular solution. We thus find that, in both cases, the sign of $\sigma(U_{0})$ determines the behavior of $\phi$ in this right neighborhood, and at sufficiently large positive (negative) $x$, the absolute value of $\phi$ may grow (decrease) enough to reach another minimum of $V_{0}$. However, $W_{\phi}$ becomes negative for sufficiently large $\phi$, which may give rise to a sign change in the right-hand side of Eq.~\eqref{Ex2}. If that happens, the field behaves as a lump in the positive half of the real line, as $\phi(x)$ returns to the same minimum it had reached at $x_0$. Thus, the matching procedure is ultimately equivalent to an initial value problem for $\delta^{\pm}(x_0)$ in $U_{0}$, with initial data that changes depending on the choice of $x_0$.

\begin{figure}[h]
	\centering
	\includegraphics[width=8cm]{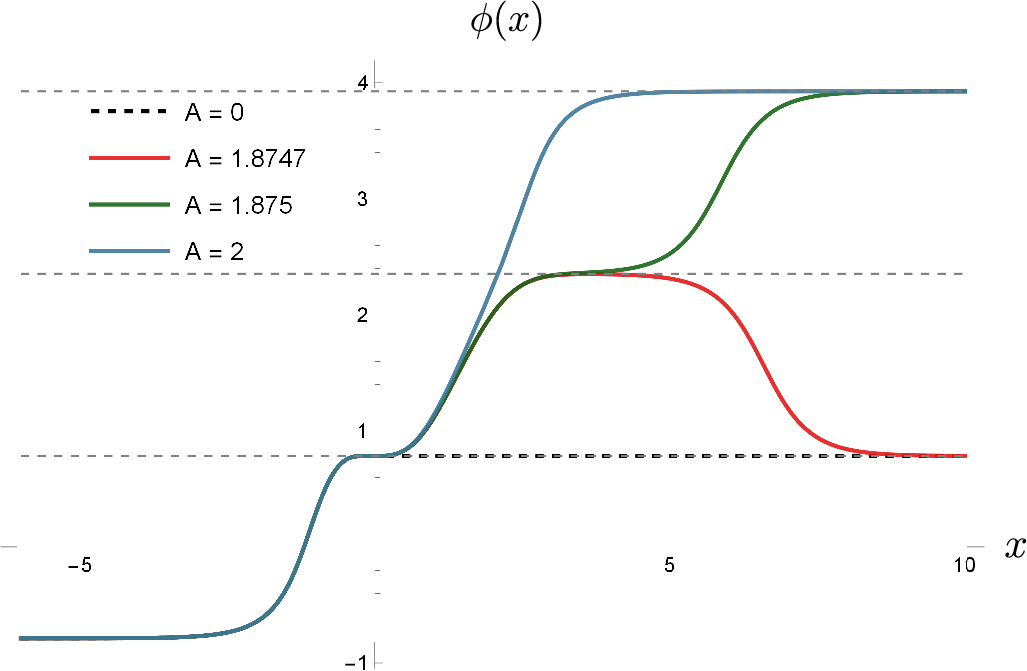}\\ \vspace{0.1cm}
	\includegraphics[width=8cm]{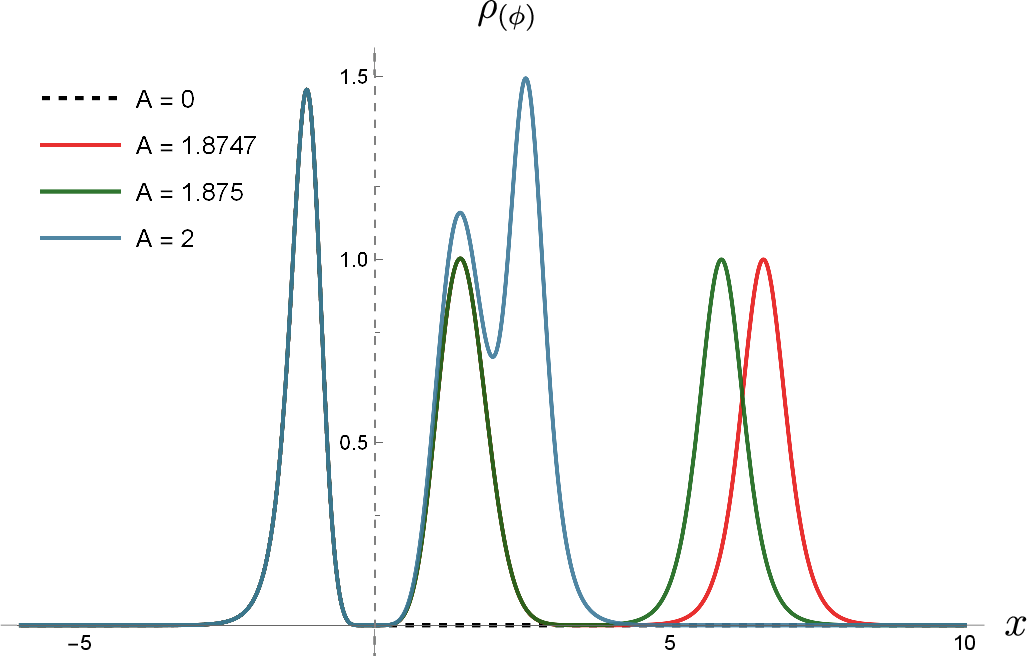}
	\caption{Solutions $\phi(x)$ of Eq,~\eqref{Ex2} (upper plot) and \qt{internal} energy density $\rho_{(\phi)}=\rho_{eff}+P\sigma^2/2$ (lower plot). Results are shown for $\alpha=1$ and three choices of $A$. Dotted grey lines represent the \qt{vacuum} values $\phi=\frac{\pi}{4},\phi=\frac{3\pi}{4}$ and $\phi=\frac{5\pi}{4}$.}
	\label{Ex2Fig}
\end{figure}

The possibilities discussed above are illustrated in Fig.~\ref{Ex2Fig}, where we depict solutions with $x_0=0$ and different values of $A$. Here we have chosen, in the $x<0$ half-line, the solution coinciding with~\eqref{Solhom1}, so that the field is nontrivial in the region where $\sigma(x)=0$ (or everywhere if $A=0$), although a configuration that is equal to~\eqref{Solhom2} inside this interval can also be found, and presents similar properties in the positive half-line. In the picture, the black dotted line represents the impurity-free scenario, where a half-compact solution is found. The blue and green lines can be interpreted as multi-kink solutions. Indeed, using the definition~\eqref{topologicalcharge} for the topological charge, we find that these configurations correspond to $N=3$ solutions in the analogue one-field theory, which according to the usual interpretation~\cite{manton} implies that the net number of kinks in the system is three. Here we have one kink interpolating between $-\pi/4$ and $\pi/4$, another from  $\pi/4$ to  $3\pi/4$ and a third one reaching $5\pi/4$ asymptotically. If the amplitude is bellow a critical value $A_{\text{crit}}$ (about $1.874758$ in this example according to our numerical computations), a solution compatible with $Q_T>1$ cannot be found for this initial data, and the field instead  returns to the previous minimum of $V_0$, presenting a lump-like behavior in the $x\geq x_0$ region. This latter situation is illustrated by the red line in Fig.~\ref{Ex2Fig}, where the kink is followed by a lump structure, qualitatively similar to kink-kink-antikink system. Although configurations with more than three kinks are not represented in the figure, we have also verified that such solutions are indeed obtained if increasingly higher amplitudes are used. Interestingly, we have not been able to find solutions representing an even number of kinks, which would amount to mixing the plus and minus branches of minima in~\eqref{vacua}. In our examples we see that the field in the positive half-line always grows from $\phi=\pi/4$ (which belongs to the \qt{plus} branch with $n=0$) towards $\phi=-\rm{arctan}(1) + \pi=3\pi/4$, which belongs to the \qt{minus} branch, but it never settles for the latter value, instead returning to the previous minimum or growing towards the next. By continuity, one could theorize that the limiting case in which $A$ is exactly equal to $A_{\text{crit}}$ would correspond to a two-kink solution with a half-compact profile.

One very important observation, and one that concerns a significant difference between the two-field systems we are dealing with and the effective theory~\eqref{effectiveLag} with metric $\tilde{g}_{\mu\nu}$, lies in the fact \textcolor{black}{that  $A_{\text{crit}}$ is \emph{not} a fixed quantity of the theory, being instead strongly dependent on the initial conditions generated by different initial kink positions.} For example, with the initial data generated by the choice $x_0=-1/2$, it can be verified that the three values of $A$ used in the figure are sufficient to obtain three-kink solutions, while some higher values of $x_0$ make it difficult to find these configurations with any choice of $A$. The relative position between $\phi$ and $\chi$ defects is thus relevant.

The above interpretations are supported by the energy density $\rho_{(\phi)}$ which we have defined in~\eqref{rhoeff} as the difference between the energy density and the coupling density $\rho_c$. As seen in the lower plot of Fig.~\ref{Ex2Fig}, this function behaves similarly to what is usually found in multi-kink systems. The green line solution has an energy density localized within three separate neighborhoods, indicating three well-separated and unambiguously defined kinks. The blue line, corresponding to the case $A=2$, lacks a plateau in the second minima, which is manifest in $\rho_{(\phi)}$ as two overlapping heaps representing a scenario in which two kinks are too close to be well-defined entities, and thus merge into a single localized structure. As is immediately seen from inspection, the energy contribution from $\rho_{(\phi)}$ increases with $|A|$, although the \emph{total} energy is the same for all the examples in the figure. Indeed, integration of $\rho_{eff}$ results, in all four cases, in a unitary contribution to the total energy $E=7/3$ of the two-field system. Although it may appear counter-intuitive that three-kink solutions have the same energy as the single kink found in the $A=0$ case, \textcolor{black}{this equivalence is explained by the fact that the absolute value of the coupling energy $-\int dx \frac{\sigma^2}{2}$ also increases with $A$. This exact compensation is typical of BPS saturation and is explained by the cancellation of dynamical forces corresponding to kink interactions and impurity coupling in the system.} For all $A\neq 0$ solutions depicted, the energy decreases by one in the $(-\frac{\pi}{4},\frac{\pi}{4})$ interval, then decreases by the same amount in $(\frac{\pi}{4},\frac{3\pi}{4})$, where $W_{\phi}$ is negative, and finally increases again in the last stretch, where the BPS contribution $\phi' W_{\phi}$ is strictly positive for both the kink and lump defects. The energy density of the $N=3$ and $N=1$ solutions become identical when the values of $A$ are similar.

We finish this work addressing the existence of multi-kink  solutions in $P=1$ theories, in order to compare these systems with the results shown above. 
When $P=1$, separable models become equivalent to a sum of two decoupled Lagrangians of the form discussed in~\cite{AdamI, AdamII}. We may therefore ignore one of the fields, say $\chi$, and treat~\eqref{FOA} as the sole BPS equation of the field theory. Let us thus consider an impurity-doped Sine-Gordon system with BPS equation
\begin{equation}\label{SineGordonImp}
	\phi'(x)=2\sin\left(\frac{\phi}{2}\right) +Ae^{-\frac{\left(x-\bar{x}\right)^2}{2}}
\end{equation}

\begin{figure}[h]
	\centering
	\includegraphics[width=8cm]{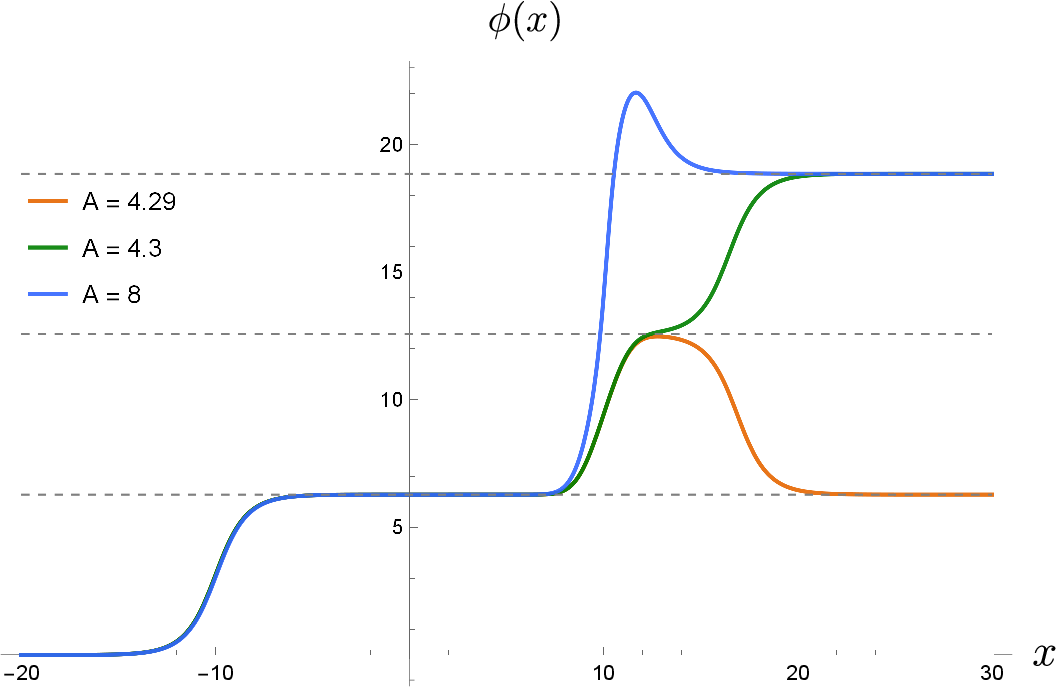}
	\caption{Solutions of Eq,~\eqref{SineGordonImp} for three values of $A$ and $\bar{x}=10$. Dotted grey lines correspond to  $\phi=2\pi$, $\phi=4\pi$ and $\phi=6\pi$.}
	\label{Ex3Fig}
\end{figure}

Now there is no impediment for the existence of a finite point $x_{0}$ such that $\phi(x_0)$ is a minimum of $V_{0}$ and, if the impurity is sufficiently strong in this region when compared to $W_{\phi}$, a multi-kink configuration may ensue according to a logic similar to the one used above. Unlike  the $P(\chi)$ model considered before, it is difficult to guess where the minima are reached, or deduce the precise analytical properties of the solution in their neighborhood. Nevertheless, it is possible to find such solutions numerically with some trial and error. To this end, the most straightforward method is to try and mimic the conditions that worked in the $P(\chi)$ model: namely construct an impurity-free solution that reaches a vacuum value at some finite point, and then force an increase of the field by adding a sufficiently strong positive-valued impurity to the vacuum region. Although solutions of the impurity-free version of the above equation do not compactify, they only differ appreciably from the minima within a finite neighborhood of the kink center. By choosing an impurity that is localized at a great distance from the original defect, we may effectively mimic the procedure used above, as there will exist a neighborhood where the equation $\phi'= \sigma(x)$ holds approximately. This has been done for Eq.~\eqref{SineGordonImp}, and solutions are depicted in Fig.~\ref{Ex3Fig}. Here we start with a kink centered at $x\approx-10$. The field is then coupled to a Gaussian impurity peaked at $\bar{x}=10$. Because $\sigma$ vanishes as $e^{-x^2}$, it is effectively zero near the left kink, so the theory is essentially homogeneous within a neighborhood of the kink center. This means that the solution is effectively identical to the usual Sine-Gordon kink in the left half of the real line. Close to $\bar{x}=10$ we find a second kink similar to the defect one would obtain by solving $\phi'=2\sin(\phi/2)+Ae^{-x^2}$ in the topological sector $(2\pi,4\pi)$. If the impurity is strong enough, the field may grow further to reach $6\pi$, giving rise to multi-kink solutions similar to those of our previous example, although here the sectors are not clearly delimitated by removable singularities. \textcolor{black}{This feature gives rise to a fundamental difference between this model and the previous one. The solutions depicted in Fig.~\ref{Ex2Fig} present a very clear separation between the first kink and the other ones. This distinction exists regardless of the distance between the first two kinks, since the singularity behaves as an infinitely massive vacuum ensuring that kinks separated by it always retain their identity. By contrast the kinks depicted in Fig.~\ref{Ex3Fig} all live in the same space and may eventually overlap.}

\textcolor{black}{As was the case in the previous example, all configurations shown in Fig.~\ref{Ex3Fig} belong to sectors with odd topological charge. This is generally true for solutions of Eqs.~\eqref{Ex2} and~\eqref{SineGordonImp} and, in fact, evaluation of $\Delta W$ shows that only odd sectors have the correct BPS energy. This feature is explained by the fact that these theories are half-BPS. This usually means that either kink or antikink sectors lose the BPS property, but for periodic potentials the situation can be different. Since the impurities are localized, the stability of boundary data is determined by the second derivatives of the superpotential. In the effective theory framework, boundary values of the kinks in Fig.~\ref{Ex2Fig} are determined by  the function ${W}_{\phi\phi}=-(1+\alpha)\sin(2\phi)$, which changes sign with shifts $\phi\to \phi + \pi/2$. For the solutions depicted in Fig.~\ref{Ex2Fig},  we have ${W}_{\phi\phi}(\pi/4)=-2$ and ${W}_{\phi\phi}(2\pi/4)=2$, thus explaining why these values act respectively as repelling and attracting fixed points. On the other hand, a BPS antikink could indeed be found between $\phi=\pi/4$ and $\phi=3\pi/4$. A similar situation is found in the Sine-Gordon models. Indeed, even in the impurity-free scenario one would need to solve the minus sign equation $\phi'=-\sin(\phi/2)$ to find a kink interpolating between $2\pi$ and $4\pi$. Thus, the adjacency between kink and antikink sectors explains why surviving BPS sectors in the half-BPS theory are not specified by the sign of the topological charge as in the $\phi^4$ model. }

\textcolor{black}{The above discussion does not preclude two-kink solutions in half-BPS theories. Indeed, the similar but non equivalent Lagrangian $\mathcal{L} = \frac{1}{2}\left[\partial_{\mu}\phi\partial^{\mu}\phi + 2\phi'\sigma(x) -\left(|W_{\phi}|+\sigma(x)\right)^2\right]$ which notably changes the coupling from $W_{\phi}\sigma$ to $|W_{\phi}|\sigma$. In this case the choice $W=-4\cos(\phi/2)$ leads to $\phi'(x)=2\left|\sin\left(\frac{\phi}{2}\right)\right| +\sigma$. We have verified that this equation can be solved for multikink configurations of any charge. The trade-offs are the absence of antikink sectors and loss of analyticity in the Euler-Lagrange equation due to the introduction of absolute values.}

%\section{Conclusion}
In this work we have introduced a model meant to generalize the framework presented in Ref.~\cite{BLM2020} in order to account for impurity coupling. We have shown a few examples of kink-like solutions and defined the precise meaning of \qt{kink} within the two-field separable theories the present work is mostly concerned with. We see that solutions can be mapped into geometrically constrained versions of the systems introduced in~\cite{AdamI, AdamII}, and have also verified that configurations consistently interpreted as multi-kink configurations are possible in this theory when regularizable poles are introduced by the function $P(\chi)$. We have also verified that multi-kink systems are possible in the models presented in~\cite{AdamI, AdamII}. These solutions are related to the ones obtained from the effective Lagrangian~\eqref{effectiveLag} in our work by the geometrical mapping $d\xi=dx/P(x)$ first introduced in~\cite{BLM2020}.

Here we have, in the interest of brevity, refrained from investigating many important aspects of the theory generated by~\eqref{Lagrangian}, such as couplings between $\sigma$ and $\chi$, stability equations, zero modes, non-separable models, and other types of solutions that may be possible in the general theory. \textcolor{black}{The solutions investigated here are stable due to BPS saturation, but the complete stability problem, which lies outside the scope of this paper, will be investigated elsewhere.} Further possible extensions involve doping of other generalized scalar models such as $K$ theories, generating fermion bound states from Yukawa-type coupling~\cite{Fermion}, and dynamical investigations such as defect collisions and scattering by impurities \cite{A1,A2,A3}.

\acknowledgments{This work is supported by the Brazilian agency Conselho Nacional de Desenvolvimento Cient\'ifico e Tecnol\'ogico (CNPq), grants Nos. 402830/2023-7 (DB and MAM), 303469/2019-6 (DB), 306151/2022-7 (MAM) and 151204/2024-1 (MAL).}


\begin{thebibliography}{99}
\bibitem{malomed}Y. S. Kivshar and B. A. Malomed, %\textit{Dynamics of solitons in nearly integrable systems},
Rev. Mod. Phys. {\bf61}, 763 (1989).
\bibitem{KinkI}F. Zhang, Y. S. Kivshar, and L. Vázquez,
Phys. Rev. A \textbf{45}, 6019 (1992).
\bibitem{KinkII}F. Zhang, Y. S. Kivshar, and L. Vázquez, Phys. Rev. A \textbf{46}, 5214 (1992).
\bibitem{KinkIII}Y. N. Gornostyrev, et al., Phys. Rev. B \textbf{71}, 094105 (2005).
\bibitem{AdamI}C. Adam and A. Wereszczynski, Phys. Rev. D \textbf{98}, 116001 (2018). 
	\bibitem{AdamII}C. Adam, T. Romanczukiewicz, and A. Wereszczynski, J. High Energy Phys. \textbf{2019}, 131 (2019). 
\bibitem{KinkKink}K. Sławińska, Phys. Rev. E \textbf{111}, 014228 (2025).
\bibitem{BLMPLB}D. Bazeia, M. A. Liao, and M. A. Marques, Phys. Lett. B, \textbf{846}, 138262 (2023).
\bibitem{BLM24}D. Bazeia, M. A. Liao, M. A. Marques, Eur. Phys. J. C \textbf{84}, 180 (2024).
\bibitem{BMM24}D. Bazeia, M. A. Marques, and R. Menezes, Eur. Phys. J. C \textbf{85}, 279 (2025).
%\textcolor{black}{\bibitem{AdamIII}C. Adam, K. Oles,  T. Romanczukiewicz, A. Wereszczynski, and  W. Zakrzewski, J. High Energ. Phys. \textbf{2021}, 147 (2021). }
\bibitem{2fieldimp} D.~Bazeia, M.~A.~Liao, and M.~A.~Marques, Chaos, Solitons \& Fractals \textbf{192}, 115950 (2025).

\bibitem{BLM2020}D. Bazeia, M.A. Liao, and M.A. Marques, Eur. Phys. J. Plus \textbf{135}, 383 (2020).

\bibitem{L1}
I. Andrade and R. Menezes, Eur. Phys. J. C \textbf{83}, 706 (2023).

\bibitem{L2}
F.C.E. Lima, R. Casana, and C.A.S. Almeida, Eur. Phys. J. C \textbf{84}, 1266 (2024). 

\bibitem{L3}M.A. Marques and R. Menezes, Chaos, Solitons and Fractals {\bf181}, 114730 (2024).
%\bibitem{L4}Adalto R. Gomes and Fabiano C. Simas, JHEP 08, 097 (2025).
\bibitem{L4}Adalto R. Gomes and Fabiano C. Simas, J. High Energy Phys. \textbf{2025}, 097 (2025).
\bibitem{L5}D. Bazeia and R. Menezes, EPL \textbf{153}, 14001 (2026).
\bibitem{L6}R. Casana, E. da Hora, and Fabiano C. Simas, Eur. Phys. J. C \textbf{86}, 125 (2026).
\bibitem{daHora} E. da Hora, L. Pereira, C. dos Santos, and F. C. Simas, Commun. Nonlinear Sci. Numer. Simul. \textbf{151}, 109070 (2025).
\bibitem{Lima} F. C. E. Lima and C. A. S. Almeida, EPL \textbf{141}, 10002 (2023).
\bibitem{rajaraman}R. Rajaraman, \textit{Solitons and instantons: an introduction to solitons and instantons in quantum field theory}, North-Holland (1984).
\bibitem{A2}C. Adam, K. Oles, T. Romanczukiewicz, A. Wereszczynski, and W.J. Zakrzewski  J. High Energy Phys. \textbf{2021}, 147 (2021).
\bibitem{bogo}E. B. Bogomol'nyi, Sov. J. Nucl. Phys. \textbf{24}, 449 (1976).




%\bibitem{ps}M. K. Prasad and C. M. Sommerfield, Phys. Rev. Lett. \textbf{35}, 760 (1975).


\bibitem{manton} N. Manton and P. Sutcliffe, Topological Solitons (Cambridge University Press, 2004). 


\bibitem{SobolevSpaces} H. Brezis, \textit{Functional analysis, Sobolev spaces and partial differential equations}, Springer (2010).

\bibitem{PDEs}L. C. Evans, \textit{Partial Differential Equations}, American Mathematical Society (2010).
\bibitem{Compactons} H. Arodź, Acta Phys. Pol. B \textbf{33},  1241 (2002).


\bibitem{Fermion}D. Bazeia and F. C. Simas, Eur. Phys. J. C \textbf{84}, 1039 (2024).

\bibitem{A1}C. Adam, K. Oles, T. Romanczukiewicz, and A. Wereszczynski, Phys. Rev. Lett. \textbf{122}, 241601 (2019).
%\bibitem{A2}C. Adam, K. Oles, T. Romanczukiewicz, A. Wereszczynski, and W.J. Zakrzewski JHEP 08, 147 (2021).

\bibitem{A3}C. Adam, N. S. Manton, K. Oles, T. Romanczukiewicz, and A. Wereszczynski, Phys. Rev. D \textbf{105}, 065012 (2022). 

	%
%	\bibitem{ps}M. K. Prasad and C. M. Sommerfield, Phys. Rev. Lett. {35}, 760 (1975).

\end{thebibliography}
\end{document}